# Three-dimensional Electromagnetic Void Space

Changqing Xu[1]†, Hongchen Chu[2]†, Jie Luo[3], Zhi Hong Hang[3],

Ying Wu[1*], Yun Lai[2*]

[1]Division of Computer, Electrical and Mathematical Science and Engineering, King Abdullah University of Science and Technology (KAUST), Thuwal 23955-6900, Saudi Arabia

[2]MOE Key Laboratory of Modern Acoustics, National Laboratory of Solid State Microstructures, School of Physics, and Collaborative Innovation Center of Advanced Microstructures, Nanjing University, Nanjing 210093, China

[3]School of Physical Science and Technology, Soochow University, Suzhou 215006, China

Emails: Ying.Wu@kaust.edu.sa, laiyun@nju.edu.cn

† These authors contributed equally.

**Abstract:**

We report a realization of three-dimensional (3D) electromagnetic void space. Despite occupying a finite volume of space, such a medium is optically equivalent to an infinitesimal point where electromagnetic waves experience no phase accumulation. The 3D void space is realized by constructing all-dielectric 3D photonic crystals such that the effective permittivity and permeability vanish simultaneously, forming a six-fold Dirac-like point with Dirac-like linear dispersions at the center of the Brillouin Zone. We demonstrate, both theoretically and experimentally, that such a 3D void space exhibits unique properties and rich functionalities absent in any other electromagnetic media, such as boundary-control transmission switching and 3D perfect wave-steering mechanisms. Especially, contrary to the photonic "doping" effect in its two-dimensional counterpart, the 3D void space exhibits an amazing property of "impurity-immunity". Our work paves a road towards the realization of 3D void space where electromagnetic waves can be manipulated in unprecedented ways.

**Introduction**

Transformation optics (TO) has provided a coordinate-transformation theory to freely manipulate waves in space and time (1-12), bestowing a rich variety of applications including negative refraction (1,2), perfect lens (3), cloaking (4,5), illusion optics (6,7) and even the simulation of geometric effects in the universe (8-10), etc. By expanding a point, a line, or a surface in free space into a finite volume of space, TO suggests this finite volume is occupied by artificial media with near-zero permittivity and permeability components. Despite of its finite volume, such an artificial medium is optically equivalent to the original geometric bodies of lower dimensions in many aspects. For instance, the expansion of a single point in free space into a three-dimensional (3D) cube with a finite volume by TO leads to 3D double-zero media (DZM) with simultaneously near-zero isotropic permittivity and permeability, i.e. $\varepsilon \approx \mu \approx 0$ (see Supplementary Materials), as shown in Fig. 1a. This region of space is similar to that of an infinitesimal point with no phase accumulation in any direction, therefore it is hereby termed as a 3D electromagnetic void space.

The realization of such a void space is closely related to zero index media (ZIM). In the past decade, prominent achievements have been made in realizing electromagnetic ZIM with zero refractive index (13-20). DZM, which exhibit good impedance-matching behaviors, have been successfully achieved at frequencies ranging from microwave to optical spectra (16-20). Nevertheless, nearly all of the previous demonstrations of electromagnetic DZM were realized in two-dimensional (2D) systems or waveguides such as optical chips and microwave parallel-plate waveguides. Up to now, 3D DZM and the corresponding effective electromagnetic void space remain unrealized and almost unexplored yet.

In this work, by using all-dielectric 3D photonic crystals (PCs) with effective double-zero indices, we demonstrate the realization of 3D electromagnetic void space. The PC is composed of a cubic lattice of dielectric meshes. When the dielectric meshes are of a certain filling ratio, the electric and magnetic dipolar states are accidentally degenerate at the Brillouin zone center, rendering a six-fold degeneracy. In the vicinity of this six-fold degenerate point, four Dirac-like linearly dispersive bands and two flat bands are formed in all directions, which is a signature of all the components of the effective permittivity and permeability simultaneously crossing zero at that frequency. The PC can thus be effectively regarded as a 3D electromagnetic void space. By fabricating such a PC at microwave frequency, we experimentally demonstrated some anomalous properties of the 3D electromagnetic void space. For instance, the profound influence of the boundary conditions of the 3D electromagnetic void space is revealed, which bestows boundary-control transmission switching and 3D perfect wave-steering mechanisms. More interestingly, we find the transmission through the space is "immune" to any impurities completely embedded in such a 3D void space, which is contrary to the photonic "doping" effect previously observed in 2D ZIM. These unusual properties, which are absent in any other electromagnetic media, clearly demonstrate that the 3D electromagnetic void space can provide a valuable and unique platform for unprecedented wave functional devices.

Figures 1b to 1e demonstrate the above-mentioned properties of the 3D electromagnetic void space formed by expanding a point to a cube via TO. First, we consider the void space bounded by two pair of plates which are, respectively, made of perfect electric conductor (PEC) and perfect magnetic conductor (PMC) in orthogonal directions, as shown in Fig. 1b. When the polarization (electric field) of the incident waves is perpendicular to the PECs, total transmission is obtained for normal incidence. When there are impurities embedded in the 3D void space, the transmission can be derived from the Maxwell equations as $T = 1 + \frac{1}{2}\left(\frac{1}{d_y H_0}\oint_{l\partial A_2}\vec{H}\cdot d\vec{l} - \frac{\omega\varepsilon_0}{d_z k_0 H_0}\oint_{l\partial A_1}\vec{E}\cdot d\vec{l}\right)$ (see Supplementary Materials), which depends on the magnetic flux through a closed curve $A1$ and the electric flux through a closed curve $A2$, as shown in Fig. 1b. When the curves $A1$ and $A2$ can be completely immersed in the 3D void space and avoid cutting through any impurities, then both the magnetic and electric fluxes are negligible due to the near-zero permittivity and permeability. This leads to total transmission that is irrespective of the material, shape, and filling ratio of the impurities. This phenomenon is amazingly contrary to the "doping effect" (21) that is observed in 2D electromagnetic or acoustic zero-index systems (21-26). Such a contrast demonstrates that the ascension of dimensionality could bring fundamentally new physics. More examples of the impurity-immunity effect are demonstrated in Supplementary Materials.

Although the transmission is almost "immune" to embedded impurities, it is very sensitive to the boundary conditions. As shown in Fig. 1c, when the PMC boundaries are replaced by PEC boundaries, the transmission is significantly reduced to near zero. This is because the wavelength within the void space is extremely large, therefore inducing the wave blocking effect of all-PEC boundaries, similar to the Faraday cage. As shown in Figs. 1d and 1e, when one PMC or PEC boundary is relocated to block the original output port, the transmission would be steered toward the directions of the removed PMC or PEC with an efficiency of 100%. This perfect wave-steering effect is also be analytically proved by using the closed curves $A1$ and $A2$ and is numerically demonstrated by using effective media (see

Supplementary Materials). In contrast, if the 3D void space is replaced by air, then the incident wave is significantly reflected backward. The controllability of the transmission by the boundaries is an amazing property of the 3D void space. In the following, we shall demonstrate the first realization of the 3D electromagnetic void space as well as its unique physical consequence with proof-of-principle microwave experiments.

Here are some additional comments. The unique new physics of 3D electromagnetic void space, such as the impurity-immunity effect, are not only essentially different from the 2D electromagnetic DZM, but also essentially different from 2D or 3D acoustic DZM. The detailed discussions and derivations about the transmission through electromagnetic/acoustic DZM with impurities in 2D and 3D space are in Supplementary Materials Note 2-5.

**Results**

**Realization of 3D electromagnetic void space by using an all-dielectric 3D PC.**

The unit cell of the 3D PC that functions as a 3D electromagnetic void space is illustrated in Fig. 2a, which has a cubic lattice with lattice constant $a$. Three identical dielectric meshes, which are orthogonally aligned along the $x$-, $y$-, and $z$-directions, intersect at the center of the unit cell, preserving the $O_h$ group symmetry. The dielectric is chosen to be aluminum oxide with relative permittivity $\varepsilon = 9$ and the host medium is air. The cross sections of the dielectric meshes are square with the side length $L = 0.32a$. The band structure of this PC along the high-symmetry lines of the first Brillouin zone (inset graph) are calculated by using COMSOL Multi-physics and plotted in Fig. 2b. At the $\Gamma$ point, we observe the accidental degeneracy of electric and magnetic dipolar states, resulting in a six-fold degeneracy. In the vicinity of the degenerate point, four bands of Dirac-like linear dispersions and two flat bands are observed. The conical and flat dispersions in the $k_x$-$k_y$ plane are plotted in Fig. 2c, which are exactly the same in the $k_y$-$k_z$ and $k_x$-$k_z$ planes. The vector-arrow field maps for the electric and magnetic dipolar states in the $z$ direction are plotted in Fig. 2d. Near the six-fold degenerate point, which is hereby denoted as Dirac-like point, the tight-binding Hamiltonian can be written in the basis of electric and magnetic dipolar states (27,28) as the following:

$$H_0 = \begin{pmatrix} E_{ex}(\vec{k}) & 0 & 0 & 0 & 2it\sin(k_z) & -2it\sin(k_y) \\ 0 & E_{ey}(\vec{k}) & 0 & -2it\sin(k_z) & 0 & 2it\sin(k_x) \\ 0 & 0 & E_{ez}(\vec{k}) & 2it\sin(k_y) & -2it\sin(k_x) & 0 \\ 0 & 2it^*\sin(k_z) & -2it^*\sin(k_y) & E_{mx}(\vec{k}) & 0 & 0 \\ -2it^*\sin(k_z) & 0 & 2it^*\sin(k_x) & 0 & E_{my}(\vec{k}) & 0 \\ 2it^*\sin(k_y) & -2it^*\sin(k_x) & 0 & 0 & 0 & E_{mz}(\vec{k}) \end{pmatrix}, \quad (1)$$

where

$$\begin{cases} E_{ei}(\vec{k}) = \varepsilon_e + 2t_e \cos(k_i) + 2t'_e \left( \cos(k_j) + \cos(k_k) \right) \\ E_{mi}(\vec{k}) = \varepsilon_m + 2t_m \cos(k_i) + 2t'_m \left( \cos(k_j) + \cos(k_k) \right) \end{cases}, \quad (2)$$

where $i, j, k \in x, y, z$, $k_x$, $k_y$ and $k_z$ are the three components of wave vector $\vec{k}$, $\varepsilon_e$ and $\varepsilon_m$ are the on-site energy, $t$, $t_e^{(')}$ and $t_m^{(')}$ represent the nearest neighbor hopping parameters. Here the lattice constant is chosen to be unit for brevity. The accidental degeneracy of electric dipolar states and magnetic dipolar states occurring at the Brillouin zone center indicates $E_e(\Gamma) = E_m(\Gamma) = E$ at the Dirac-like frequency $\omega_\Gamma = c\sqrt{E}$, the eigenvalues of the Hamiltonian are obtained as

$$\begin{cases} E_1 = E_2 = E \\ E_3 = E_4 = E + 2|t|\sqrt{\sin^2(k_x) + \sin^2(k_y) + \sin^2(k_z)}, \\ E_5 = E_6 = E - 2|t|\sqrt{\sin^2(k_x) + \sin^2(k_y) + \sin^2(k_z)} \end{cases} \quad (3)$$

Given that $\sin(i) \approx i$ when $i$ approaches zero, and $k = \sqrt{k_x^2 + k_y^2 + k_z^2}$, we obtain

$$\begin{cases} E_1 = E_2 = E \\ E_3 = E_4 = E + 2|t|k \\ E_5 = E_6 = E - 2|t|k \end{cases}. \quad (4)$$

The dispersion near the Dirac-like point can be obtained as

$$\begin{cases} \omega_1 = \omega_2 = \omega_\Gamma \\ \omega_3 = \omega_4 = \omega_\Gamma + \dfrac{2|t|kc^2}{\omega_\Gamma} \\ \omega_5 = \omega_6 = \omega_\Gamma - \dfrac{2|t|kc^2}{\omega_\Gamma} \end{cases} \qquad (5)$$

which indicates two flat bands ($\omega_1$ and $\omega_2$) and four linear bands ($\omega_3$ to $\omega_6$) near the $\Gamma$ point.

The coexistence of Dirac-like linear dispersions and flat dispersions near the Dirac-like point can also be understood from an effective medium point of view. When the effective permittivity and permeability cross zero at the same frequency $f_{Dirac}$, Dirac-like linear dispersions are naturally formed (see Supplementary Materials). The flat bands corresponds to the longitudinal electric or magnetic dipolar modes, which emerge when the effective permittivity $\varepsilon_{eff}$ or permeability $\mu_{eff}$ crosses zero (see Supplementary Materials), respectively. Here we emphasize that such longitudinal dipolar modes near the $\Gamma$ point are almost impossible to be excited, because incident electromagnetic waves in free space are transverse waves. Based on an effective parameter retrieval method (29-31), we have calculated the $\varepsilon_{eff}$ and $\mu_{eff}$ of the PC, as plotted in Fig. 2e. Indeed, at the frequency of the Dirac-like point $f_{Dirac} = 0.54c/a$, both $\varepsilon_{eff}$ and $\mu_{eff}$ cross zero simultaneously. This is consistent with the theory about the relationship between zero effective parameters and Dirac-like linear dispersions as well as flat dispersions.

We fabricate the above 3D PC with a lattice constant $a = 1.5cm$. The frequency of the Dirac-like point is $f_{Dirac} = 10.9GHz$. We measure the transmission spectrum through this PC at different angles to verify the band dispersions. The experimental setup is shown in Fig. 3a. The PC is composed by $4 \times 10 \times 20$ units of aluminum oxide meshes and placed on a rotating platform. An antenna horn is placed behind a perforated sponge layer with a hole of the diameter 13cm, which generates a Gaussian beam to illuminate on the PC. A probe is placed behind the PC to detect the transmission. Surrounding sponge layers can eliminate the reflection from the environment. Figures 3c and 3e show the experimentally measured transmittance through the PC under the incidence of transverse electric (TE) and transverse magnetic (TM) polarizations, respectively. The corresponding numerical results are shown in Figs. 3b and 3d, which agree well with the experimental results. A cone of high transmission is observed, which resembles the band structure with the vertex appearing at the Dirac-like point around 10.9GHz. Outside of the cone, low transmission is observed. It should be noted that the transmission at the vertex of cone is not the highest. This is because that the PC, as an effective DZM, would radiate waves outward from each surface of the PC at $f_{Dirac}$, reducing the transmission detected by the probe. The flat bands near the Brillouin zone center are not excited due to their physical nature of longitudinal dipolar modes. Only when far away from the Brillouin zone center, where the local effective medium approximation breaks down, the bands are no longer flat and deviate from longitudinal waves, leading to high transmission. Overall, the properties of the 3D electromagnetic void space are well maintained near the Dirac-like point at the Brillouin zone center.

**Unique physics of the 3D electromagnetic void space.**

As mentioned earlier, the controllability of the boundaries on transmission is one of the unique physics of the 3D void space. This leads to boundary-control transmission switching and 3D perfect wave-steering mechanisms that are absent in any other electromagnetic media. In the following, a proof-of-principle experimental verification is performed. The experimental setup is illustrated in Figs. 4a and 4b, in which the PC is composed of $10\times10\times10$ unit cells. For a waveguide configuration, we cover the PC by PEC and PMC plates in orthogonal directions. In the microwave spectrum, the PEC plates are simply metal plates, while the PMC plates can be approximated by using metal plates with an additional distance of $\lambda/4$ from the surface of the PC sample (32). This approximation has been demonstrated to be accurate when the PC functions as an effective DZM in the long wavelength regime (32). By adjusting the distance between the metal plates and the PC, the effective switch between PEC and PMC plates can be approximated. First, we cover the top and bottom surfaces of the PC with PEC plates, and the left and right sides with PMC plates. We measured the distribution of $E_z$ in the transmission domain for a TE wave incidence along the $x$ direction at $f_{Dirac}=10.9GHz$. The data are obtained by scanning the region (purple square) behind the PC, as shown in Fig. 4. Both simulated and experimental results show high transmission with almost planar wavefront, as plotted in Figs. 4c and 4d, respectively. This results from the formation of transverse electric magnetic (TEM) waveguide by the PEC/PMC plates as well as the impedance matching of DZM to the background (16-20, 33-35). However, when the effective PMCs are replaced by PECs via removing the $\lambda/4$ distance between the PC and the metal plate, both simulated and experimental results indicate very low transmission through the PC, as shown in Figs. 4e and 4f. This transmission switching phenomenon can be explained by the wave blocking effect of the Faraday cage (metallic meshes), which appears when the wavelength is much larger than the scale of the meshes. At $f_{Dirac}$, the PC serves as 3D electromagnetic void space and the effective wavelength approaches infinity, therefore enabling the wave blocking effect. Such a boundary-control transmission switching mechanism is a unique feature of the 3D electromagnetic void space. Without the PC as the 3D void space, the metal waveguide alone cannot block the wave because the wavelength in free space (2.75cm) is much smaller than the size of the PC (15cm).

The 3D electromagnetic void space also enables 3D wave steering to other directions with high efficiency. As illustrated in Fig. 5a, one of the PMC plate in the *x-z* plane is moved to the back of the PC in the *y-z* plane. Such a change efficiently steers the incident waves into the direction where the PMC plate was originally located. Through numerical simulation and experimental measurement, we have obtained and plotted the distributions of $E_z$ in the output domain (marked by purple in Fig. 5a) in Figs. 5b and 5c, respectively. These results evidently show high transmission with almost planar wave front, which prove that high-efficiency steering effect. This phenomenon can be intuitively understood as a consequence of the optically "void" characteristic, which effectively pulls the incident surface and the exit surface together. For comparison, we have also investigated the case of the same waveguide configurations without the PC. The steering effect disappears and the transmission is significantly reduced, as shown in Figs. 5d and 5e, indicating most of the incident waves are reflected back in the *y-z* plane. In Fig. 5f, we consider another case where one PEC plate on the top is moved to the back of the PC in the *y-z* plane. The simulated and experimentally measured electric field distributions are shown in Figs. 5g and 5h, manifesting the wave steering effect. The horizontal incident wave is redirected to the vertical direction, preserving the wavefront and amplitude nicely. The outgoing wave is converted to the $E_x$ polarization due to the transverse nature of electromagnetic wave. In contrast, without the PC, the transmission is extremely low due to the backward reflection, as shown in Figs. 5i and 5j. Such a 3D perfect wave steering mechanism proves the validity of PC as the 3D electromagnetic void space (see Supplementary Materials).

Besides the amazing phenomenon demonstrated above, one fundamental physical difference (23,27) between the 2D and 3D DZM lies in their responses to embedded impurities. In two dimensions, the impurities can dramatically modify the bulk properties of 2D ZIM, and tune the transmission behavior from total reflection to total transmission (21-26). Such an effect is denoted as the photonic "doping" effect (21). An intuitive thinking may lead to a conjecture that this "doping" effect may also work for 3D ZIMs. However, theoretical analysis shows that wave propagation in 3D electromagnetic DZM is almost "immune" to the embedded finite-sized impurities, which is contrary to the case of 2D ZIM (33). Total transmission is guaranteed, regardless of the geometries, material properties and size of the finite-sized impurities. The details of derivations can be found in Supplementary Materials. Such an impurity-immunity effect remains experimentally unexplored due to the lack of 3D electromagnetic DZM. In what follows, we shall demonstrate this counter-intuitive phenomenon.

The experimental setups for measuring the forward scattering effect of a metamaterial scatterer without and with the PC are shown in Figs. 6a and 6b, respectively. The metamaterial scatterer is set to be a polarization converter with a size of $2a \times 2a \times 2a$. It can efficiently transform the polarization of incident waves from $E_z$ to $E_y$. The details of the metamaterial polarization converter can be found Supplementary Materials. An incident beam with $E_z$ polarization is incident along the *x*-direction. Through numerical simulation and experimental measurement, we have obtained and plotted the distributions of $E_y$ and $E_z$ in a region behind the waveguide (the dashed box) in Figs. 6c to 6h for cases without and with the presence of PC, respectively. When the PC is removed and only the bare metamaterial scatterer remains, a large portion of wave is converted from the $E_z$ polarization to the $E_y$ polarization in the outgoing wave. However, when the metamaterial scatterer is embedded in the PC as the 3D electromagnetic void space, the $E_y$ component in the outgoing wave disappears, indicating the scattering effect is evidently "disabled" by the void space. This proof-of-principle experiment is a strong evidence of the impurity-immunity effect of the 3D electromagnetic void space (33,36). More demonstrations of the immune effect of impurities with different geometries and material properties can be found in the Supplementary Materials.

**Discussions and Conclusions**

As demonstrated above, the 3D electromagnetic void space exhibits unique physical properties that do not exist in any other electromagnetic media. In normal electromagnetic media, the wavelength is always finite and the influence of the boundary conditions on transmission is reduced to a negligible level at places far enough away from the boundary. In a 3D void space, the wavelength approaches infinity, therefore its boundary conditions always have a dramatic influence on the transmission. For the same reason, when the impurities are completely embedded in the void space, the scattering phenomenon is reduced significantly when the wavelength approaches infinity, enabling the impurity-immunity effect. For a random distribution of impurities, percolation-like phenomenon of light, i.e. squeezing through the gaps between impurities like water, are anticipated (33).

The design principle of the void space via the PC is general. By adjusting the permittivity of the dielectric meshes and the filling ratio accordingly, the 3D electromagnetic void space with unit cells at a deep sub-wavelength scale can be fabricated in principle. In a 3D PC with $O_h$ group symmetry, the accidental degeneracy between the electric and magnetic dipolar states is necessary for the creation of the Dirac-like conical dispersions with additional flat dispersions. The effective permittivity and permeability of the PC simultaneously cross zero at the frequency of Dirac-like point. The 3D electromagnetic void space provides a platform for abundant physics and novel phenomena, such as the boundary-control transmission switching, 3D perfect wave steering, and impurity-immunity effect for arbitrary impurities. Our demonstration of the above unique physical properties not only proves the validity of the 3D electromagnetic void space, but also paves a road for unprecedented wave-functional devices.

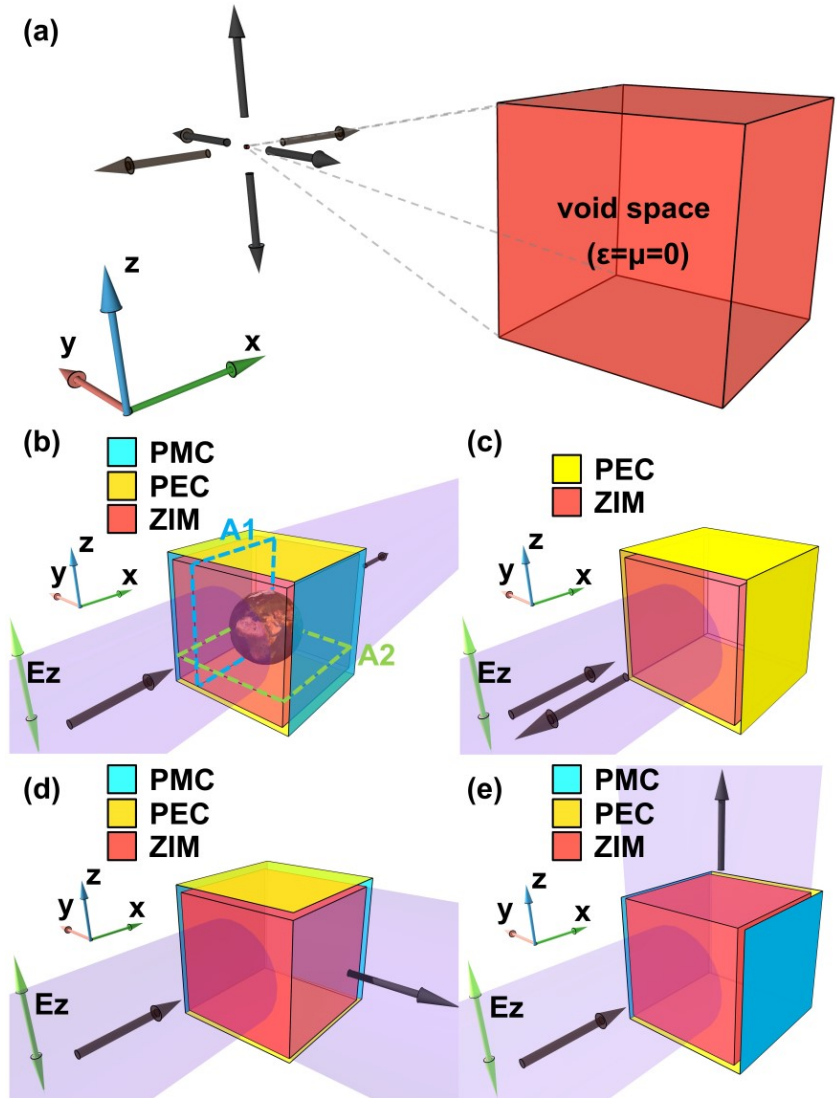

Fig. 1 Unique properties of the 3D electromagnetic void space. (a) The expansion of an infinitesimal point of space into a cube of a double-zero medium (DZM). (b) The transmission through such an optical void space is almost "immune" to any impurities which completely embedded in DZM. The perfect electric conductor (PEC) walls (colored yellow) and perfect magnetic conductor (PMC) walls (colored blue) bounds to the surface of DZM, builds up a transverse electric magnetic waveguide. The incident beam with $E_z$ polarization comes from positive *x*-direction. (c) The DZM is in a metallic waveguide when the PMC walls in (b) are replaced by PEC walls. The incident wave is significantly reflected by metallic walls due to the infinite wavelength in DZM. When one PMC or PEC in (b) is relocated to block the original output port, the transmission can be flexibly steered toward the (d) negative *y*-direction and (e) positive *z*-direction.

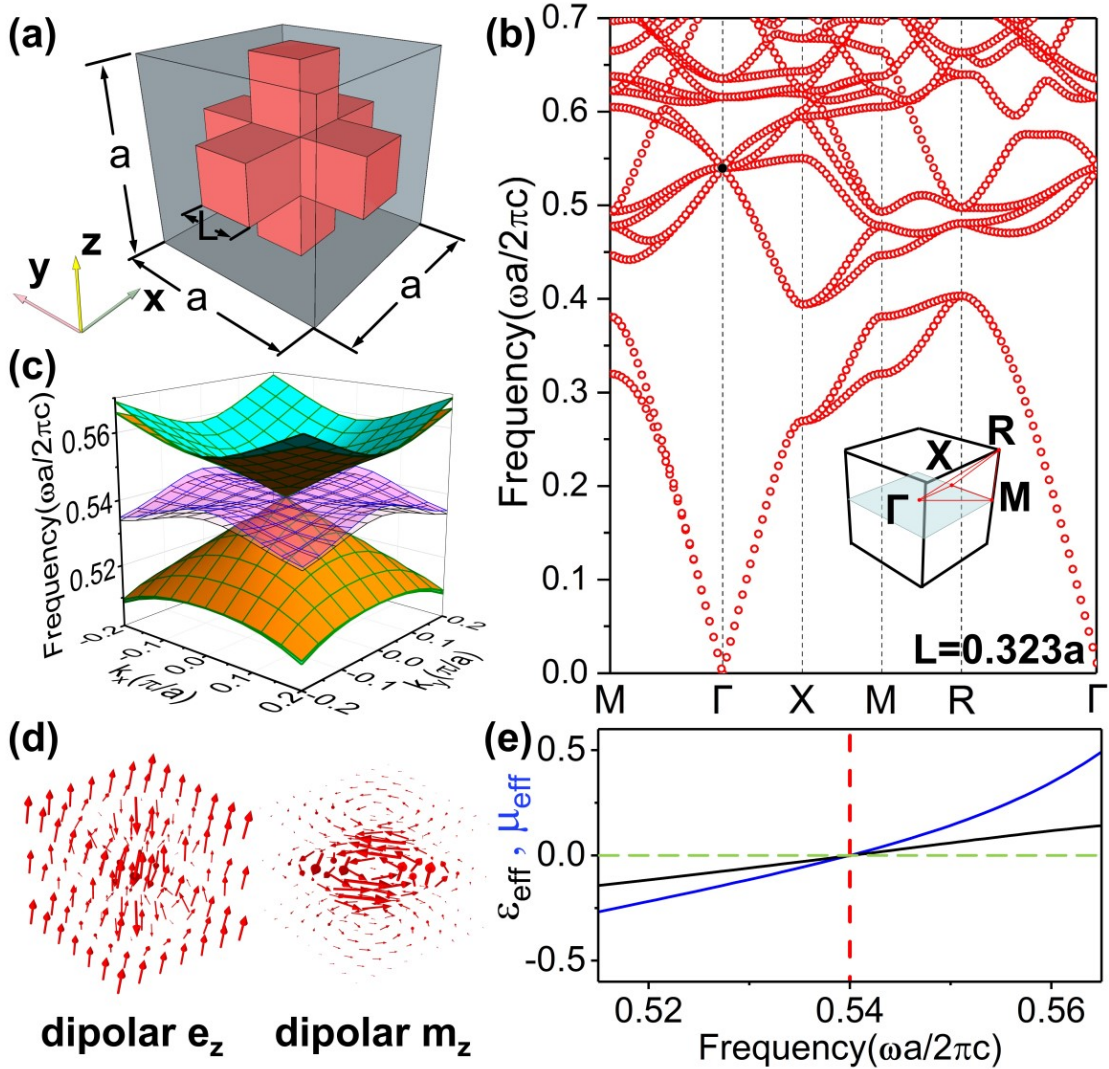

Fig. 2 Realization of 3D electromagnetic DZM in a photonic crystal. (a) The unit cell of the 3D PC with lattice constant *a*. The host medium (gray) is air. The dielectric meshes (red) are aluminum oxide blocks ($\varepsilon = 9$) with side length $L \times L \times a$. (b) Band structure along the high-symmetry lines with $L = 0.32a$, which shows a six-fold Dirac-like point (the black dot) with conical dispersions in its vicinity at the Γ point. (c) The conical dispersion surfaces near the Dirac-like point in the $k_x - k_y$ plane. (d) Arrow maps of the electric field, exhibits electric "dipolar z" states and the magnetic "dipolar z" states. (e) The relative permittivity (black curve) and relative permeability (blue curve) from the effective parameter retrieval method. The red dashed line corresponds the frequency of Dirac-like point.

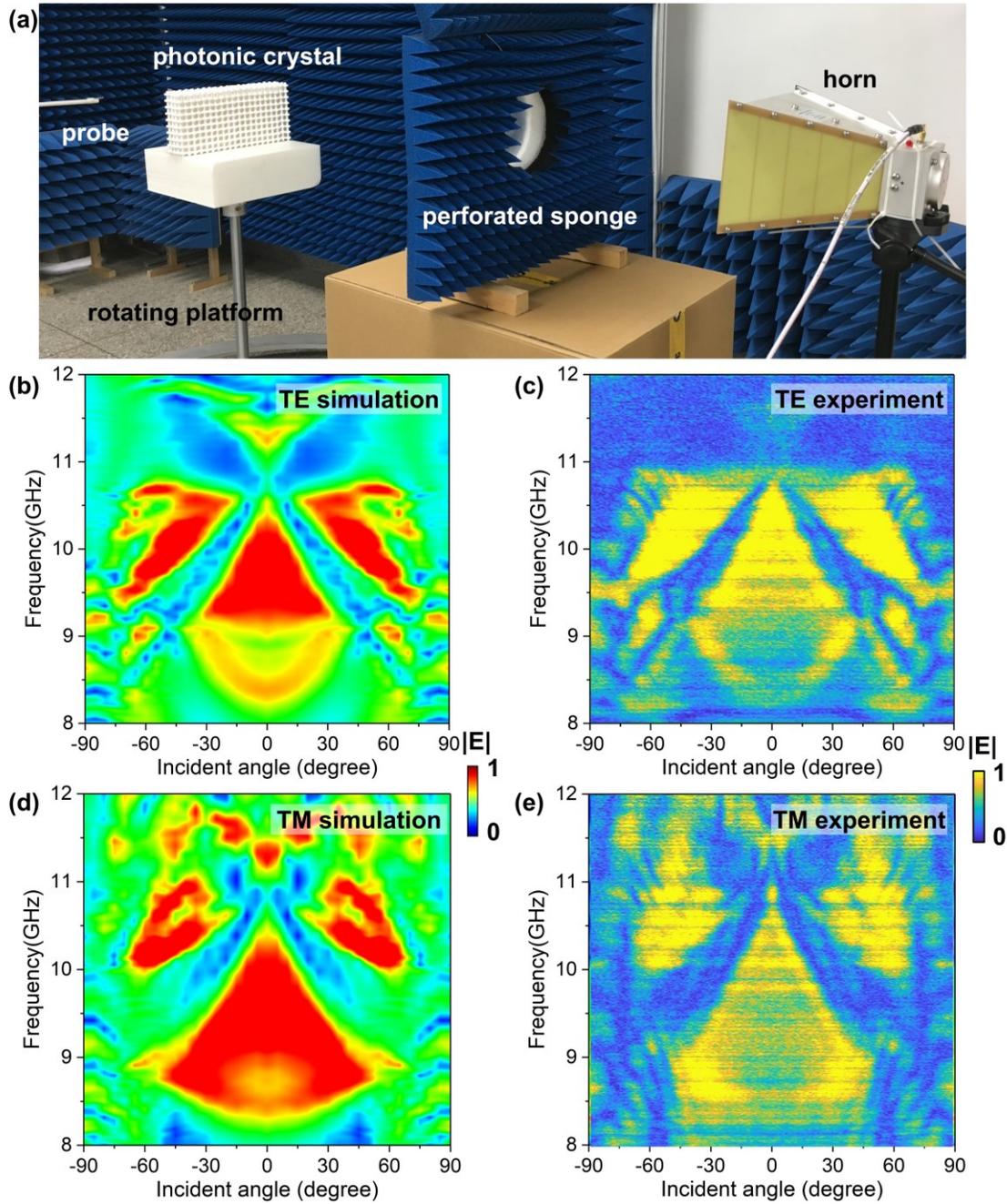

Fig. 3. The transmission spectrum near the frequency of Dirac point $f_{Dirac}=10.9GHz$ with the lattice constant $a=1.5cm$. (a) Experimental setup. (b) Numerical and (c) experimental results from a probe behind PC for transverse electric (TE) incidence. (d) Numerical and (e) experimental results for transverse magnetic (TM) incidence.

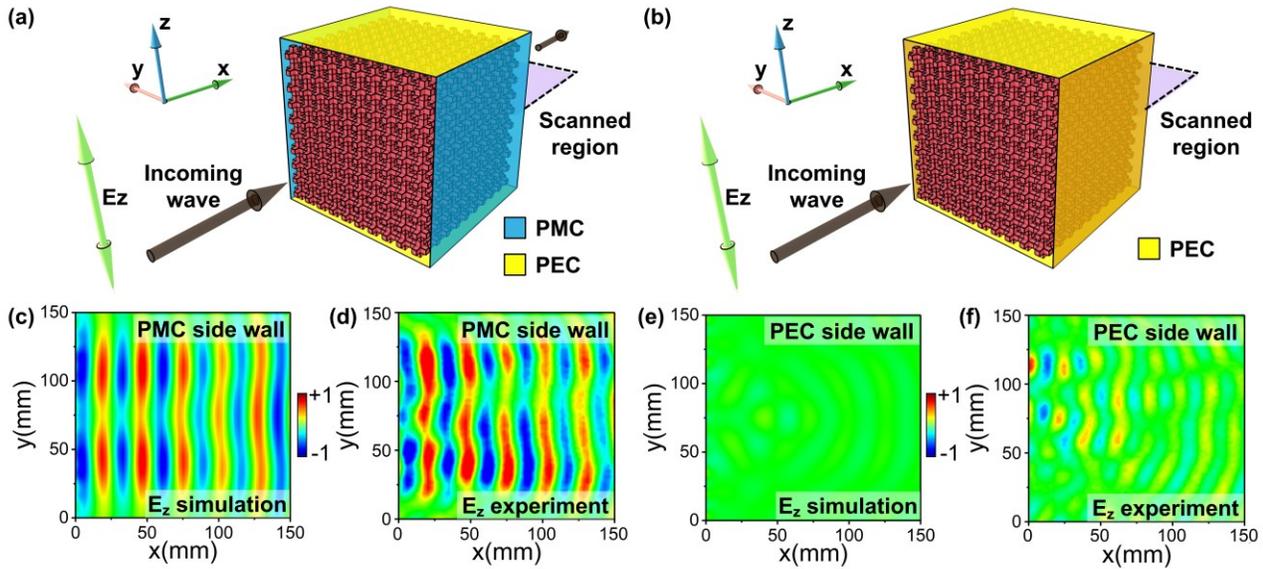

Fig. 4. The boundary-controlled transmission switching effect. (a,b) Illustration of the boundary-controlled transmission switching effect with a PC composed of 10×10×10 unit cells. The top and bottom sides of PC attach two PEC walls (colored yellow), the left and right sides in *x-z* plane can be switched between (a) PMC (colored blue) and (b) PEC. TE wave comes from the positive *x* direction. (c) Simulated and (d) experimental electric field distributions in the region colored purple in (a). (e) Simulated and (f) experimental results in the region colored purple in (b).

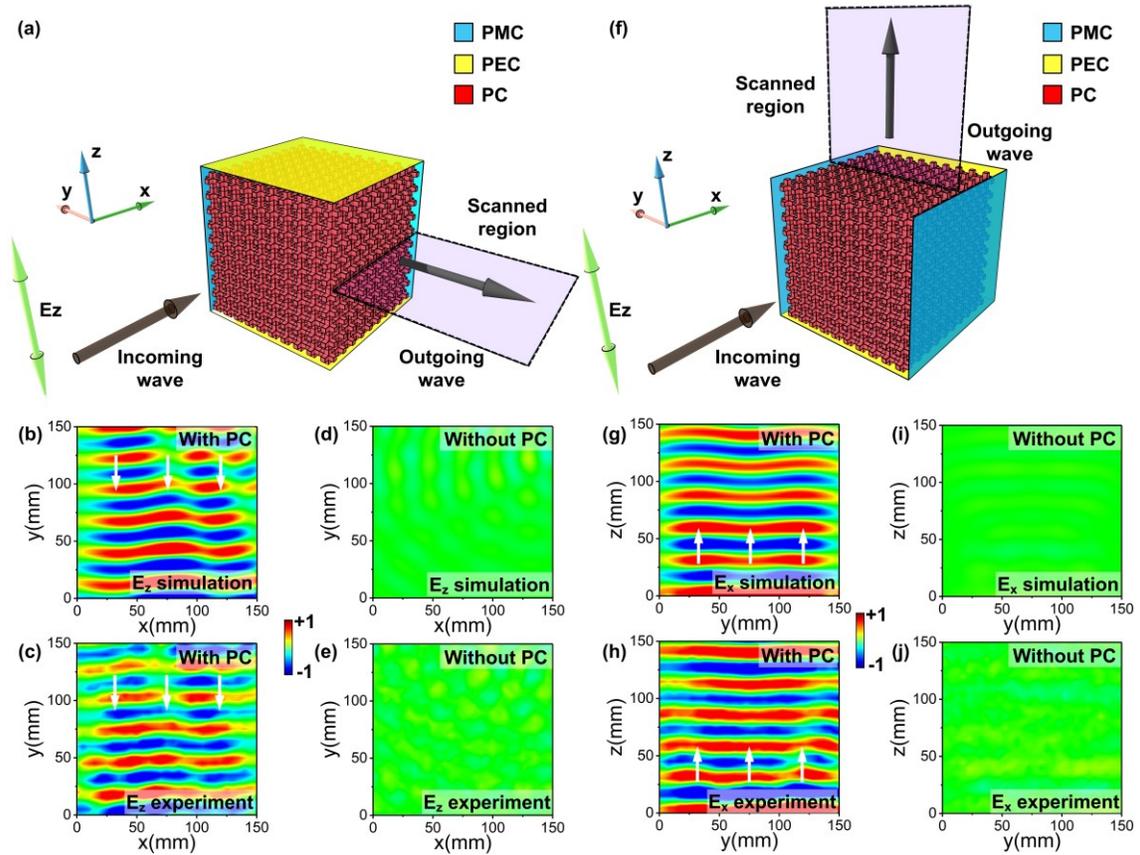

Fig. 5. Perfect wave steering in 3D space. (a) Illustration of the wave steering, where the PMC attached to the right side of the void space in Figure 4(a) is relocated to block the original output port of transmission. At the frequency of the Dirac-like point, a TE incident beam comes from positive $x$-direction. (b) Simulated and (c) experimental results of electric field in the region marked by purple in (a). (d) Simulated and (e) experimental results of electric field in the purple region when the PC is removed. (f) Illustration of another case of wave steering. The PEC attached to the top of the void space in Figure 4(a) is relocated to block the original output port. (g) Simulated and (h) experimental results of electric field distribution $E_x$ in the region marked by purple in (f). (i) Simulated and (j) experimental results of electric field in the purple region in (f), but without the PC.

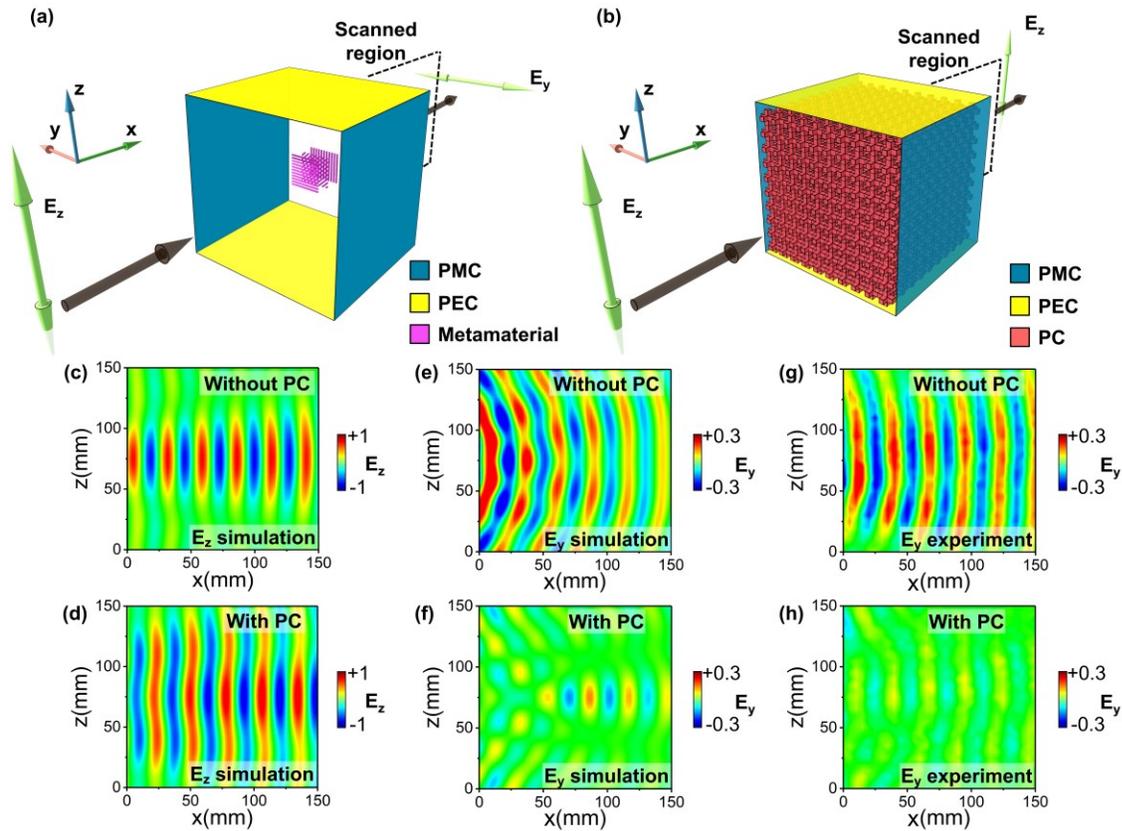

Fig. 6. The forward scattering effect of a metamaterial polarization converter without and with the PC. (a) Illustration of the metamaterial polarization converter in a transverse electromagnetic waveguide. The incident beam with $E_z$ polarization comes along positive *x*-direction. (b) The region between metamaterial and waveguide in (a) is filled with our PC. (c,d) The simulated $E_z$ field in the dashed box in (a) and (b). (e,f) The simulated $E_y$ field in the dashed box in (a) and (b). (g,h) The experimental results of $E_y$ field in the dashed box in (a) and (b).